\documentstyle[ws-procs975x65,epsfig]{article}

\def\be{\begin{equation}}
\def\ee{\end{equation}}

\begin{document}

\title{TIME AND SPACE VARIATION OF FUNDAMENTAL CONSTANTS: MOTIVATION AND LABORATORY SEARCH
}


\author{Savely G. Karshenboim}

\address{D. I. Mendeleev Institute for Metrology, 198005 St. Petersburg, Russia\\
Max-Planck-Institut f\"ur Quantenoptik, 85748 Garching, Germany\\ E-mail: sek@mpq.mpg.de}   


\maketitle
\abstracts{ 
Fundamental physical constants play important role in modern physics. 
Studies of their variation can open an interface to new physics. An 
overview of different approaches to a search for such variations is 
presented as well as possible reasons for the variations. Special 
attention is paid to laboratory searches.}

\section{Introduction}

Any interactions of particles and compound objects such as atoms and 
molecules are described by some Lagrangian (Hamiltonian) and {\em constancy} 
of parameters of the basic Lagrangian is a cornerstone of modern physics. 
Electric charge, mass and magnetic moment of the particle are  
parameters of the Lagrangian. However, there are a few simple reasons 
why we have to expect the nature to be not so simple.
\begin{itemize}
\item A theory described by a Lagrangian suggests some properties 
of the space-time. It seems that introducing gravitation we arrive 
to some inconsistency of a classical description of the space-time 
continuum and that means that the picture must be more complicated. It 
is not necessary, however, that the complicated nature imply variable 
constants, but it is possible.
\item In particle/nuclear/atomic/molecular physics we deal with
the effective Lagrangians. The ``true'' fundamental Lagrangian 
is defined at the Planck scale for elementary objects (leptons, 
quarks and gauge bosons) and we can study only its ``low-energy'' limit 
with a pointlike electron and photon and extended hadrons and nuclei. 
\item One more reason is presence of some amount of matter, which selects 
a preferred frame and produces some background fields. In usual 
experiments we often have problems with environment and have either 
to produce some shielding or to subtract the environment contribution. 
However, we cannot ignore the whole Universe and its evolution.
\item The expansion of Universe may lead to some specific time and 
space dependence in atomic transitions which are similar to a variation 
of ``constants''.
\end{itemize}
An illustration  can be found in the so-called inflation model 
of evolution of the Universe (see e.g. \cite{sgk:inflation}). 
The Standard Model of  evolution suggests a phase transition in some 
very early part which dramatically changed properties of electrons 
and photons. It happens without any changes of the unperturbed parameters 
of the basic Lagrangian defined at the Planck scale. A change of the 
electron mass (from zero to some nonvanishing value of $m_e$) 
arose eventually from cooling of matter caused by expansion. Meanwhile 
photon properties were changed via renormalization going from the 
Planck scale down to our real world (which is very different for a 
zero and non-zero electron mass).

Considering variation of the fundamental constants we have to 
clearly recognize two kinds of a search. The first one is related 
to the most sensitive and  easily accessible quantities. In such 
a case a limitation for the variation is the strongest and easiest 
to obtain, but sometimes it is not clear what fundamental quantity 
it is related to. An example is a study of samarium resonance by 
absorption of a thermal neutron \cite{sgk:shlyakhter}
\be\label{sgk:149}
^{149}{\rm Sm} + n \to {}^{150}{\rm Sm} + \gamma\;.
\ee
Estimations led to an extremely low possible variation but it is hard to 
express it in terms of the fine structure constant or some other 
fundamental constant (see Sec. \ref{sgk:sum} for detail). 

The other kind of a search is provided by a study of quantities which can 
be clearly related to the fundamental constants such as optical 
transitions (see \cite{sgk:60} and Sec. \ref{sgk:opt} for detail). 

One may wonder whether it is really important to interpret a variation 
of some not fundamental value (such as a position of a resonance) 
in terms of some fundamental quantities. A fact of the variation 
itself must be a great discovery more important than the exact value 
of the variation rate of the fine structure constant $\alpha$ or 
another basic constant. A problem, however, is related to the nature 
of precision tests and searches. Any of them is realized on the edge 
of our ability to perform calculations and measurements and any single 
result on the variation is not sufficient since a number of sources 
of new systematic effects, which were not important previously at
the lower level of accuracy, may appear now. It is crucially important 
to be able to make a comparison of different results and to check 
if they are consistent.

In our paper we first try to answer a few basic questions 
about the constants:
\begin{itemize}
\item Are the fundamental constants fundamental?
\item Are the fundamental constants constant?
\item What hierarchy of the variation rate can be expected for 
various basic constants?
\end{itemize}

After a brief overview of most important results we consider 
advantages and disadvantages of laboratory searches and in particular 
experiments with optical frequency standards. 

\section{Are the fundamental constants fundamental?}

First of all, we have to note that we are mainly interested in searches 
for a possible variation of dimensionless quantities. A search of the 
variation of constants is based on comparison of two measurements of the 
same quantity separated in time and/or space. For such a comparison, 
the units could also vary with time and their realization should be 
consistent for separate measurements. In principle, we can compensate 
or emulate a variation of a dimensional quantity via a redefinition of 
the units. To avoid the problem we have to compare dimensionless 
quantities, which are {\em unit}-independent. E.g., studying some 
spectrum we can make a statement on the variation of the fine structure 
constants $\alpha$, but not on the variation of speed of light $c$, 
Planck constant $\hbar$ or electric charge of the electron $e$ separately.

However, the variation of dimensional quantities can in principle be 
detected in a different kind of experiment. If we have to recognize 
which constant is actually varying, we should study effects due to 
their time- and space- gradients. We do not consider such experiments 
in this paper.

Precision studies related to astrophysics as well as atomic and 
nuclear physics deal with characteristics which can be linked to the 
values of the charge, mass and magnetic moment of an electron, proton 
and neutron, defined as their properties for real particles 
(i.e. at $E^2=p^2c^2+m^2c^4$) at zero momentum transfer. In the case of 
nuclear transitions, variation of the pulsar periods {\em etc} we can 
hardly interpret any results in terms of the fundamental constants, 
while in the case of atomic and molecular transitions that can be done 
(see Sec. \ref{sgk:trans}). 

We can combine the constants important for spectroscopy into a small
number of basic combinations:
\begin{itemize}
\item one dimensional constant (e.g., the Rydberg constant $Ry$) is needed 
to describe any frequency;
\item a few basic dimensionless constants, such as
\begin{itemize}
\item the fine structure constant $\alpha$; 
\item the electron-to-proton mass ratio $m_e/m_p$;
\item the proton $g$ factor $g_p$;
\item the neutron $g$ factor $g_n$
\end{itemize}
are needed to describe any ratio of two frequencies.
\end{itemize}

As mentioned above, any variation of a dimensional constant cannot 
be successfully detected: in the case of the astrophysical measurement 
it will be interpreted as a contribution to the red shift and 
removed from further analysis, while in the laboratory experiments 
it will lead to the variation of the second, defined via cesium 
hyperfine structure. A variation of the value of the Rydberg constant 
in respect to the cesium hyperfine interval is detectable since it is 
a dimensionless quantity. However, a physical meaning of such variation 
can not be interpreted in terms of the Rydberg constant as a fundamental 
constant, its possible variation should be due to a variation of the 
cesium magnetic moment (in units of the Bohr magneton) and the fine 
structure constant.

Nature of the $g$ factor of the proton and neutron is not well 
understood and in particular it is not clear if their variations can 
be considered as independent. Obviously, the $g$ factors are 
{\em not} truly fundamental constants, arising as a result of strong 
interaction in the regime of strong coupling. 

Concerning the fine structure constant, we first have to mention that 
it is a result of renormalization while some more fundamental quantities 
are defined at the Planck scale. 

The origin of the electron and proton mass is different. The electron 
mass is determined by the details of the Higgs sector of the Standard Model 
of the electroweak interactions, however, this sector originates from 
some higher-level theory and a really fundamental constant is 
rather $m_0/M$, where $m_0$ is a ``bare'' electron mass (i.e. the mass 
prior to the renormalization which is needed to reach the electron 
mass $m$ for a {\em real} electron) and $M$ is a ``big'' mass related to 
some combination of the Planck mass and the compactification radius 
(if we happen to live in a multidimensional world). In the case of proton 
the situation is different. Most of the proton mass is proportional to 
$\Lambda_{\rm QCD}$ (see e.g. \cite{sgk:marciano,sgk:fritzsch}), 
which can be expressed in terms of the unperturbed interaction constant 
and a big mass $M^\prime$. The latter is some combination of the Planck 
mass and compactification radius, but it is not the same as $M$. A 
small portion of the proton mass and in particular $m_p-m_n$ comes from 
the mass of current quarks, theory of which is similar to theory of the 
electron mass. The values of $m_0/M$ and $\Lambda_{\rm QCD}/M$ can
in principle be expressed in terms of the parameters of the basic Lagrangian 
defined at the Planck scale. 

Studies of the gravitational interaction can provide us with a limitation 
for a variation of $m_p/M$, however, the limitations are much weaker 
than those obtained from spectroscopy (see e.g. \cite{uzan}. Performing spectroscopic 
measurements we can reach a limitation for a value of $m_e/m_p$, 
however, it is rather an accidental value, in contrast to 
$m_0/M$ and $\Lambda_{\rm QCD}/M$, and its interpretation involves 
a number of very different effects.

\section{Are  fundamental constants constant?}

We have to acknowledge that some variations, or effects which 
may be interpreted as variations have happened in past and are present now.
\begin{itemize}
\item A Standard model of the evolution of our Universe, has a special 
period with {\em inflation of Universe} due to a phase transition which 
happened at a very early stage of the evolution and significantly 
changed several properties of particles (see e.g. \cite{sgk:inflation}). 
In particular, the electron mass and so-called current quark masses 
(the latter are responsible for a small part of the nucleon mass and 
in particular for the difference of the proton and neutron mass) were 
changed. Prior to the phase transition the electron was massless. The 
proton mass determined by so called $\Lambda_{\rm QCD}$ was essentially 
the same. At the present time the renormalization of the electric charge 
only slightly affects the charge because it has an order of 
$\alpha/\pi \ln(M/m)$. However, with massless leptons the renormalization 
has not only ultraviolet divergence but also an infrared one. The phase 
transition for the electron mass $m$ is also a phase transition for its 
electric charge $e$. The transition was caused by cooling of the Universe, 
and cooling was a result of expansion. The Universe is still 
expanding and cooling. It should lead to some variation of $m$ and $e$ 
but significantly below a level of accuracy available for 
experiments and observations now.
\item Expansion of the Universe should modify the Dirac equation for 
the hydrogen atom and any other atoms and nuclei. However, the expansion 
itself, without any time and space gradients will just create 
a red shift common for any atoms and transitions in an area. The second 
order effect gives an acceleration (note that for a preliminary estimation 
one can set $H^\prime \sim H^2$). The acceleration will shift energy 
levels but produce no time variation. And only the $H^3$ term can give 
a time dependent correction to the energy levels. It is indeed beyond 
any experimental possibility.
\item In principle, we also have to acknowledge that if the Universe 
has a finite size, that must produce an infrared cut off which should 
enter into equations. Since we do not have any real infrared divergence 
for any observable quantity, the radius of the Universe will enter 
the expressions for the electric charge and mass of electron in 
combinations such as $(a_0/R_U)^2$ and the ratio of the Bohr radius and 
the radius of the Universe is extremely small. With the expansion of 
the Universe, the radius $R_U(t)$ is time dependent and that will give 
some small (undetectable) variation of the constants. The real 
situation is not so simple. First, we do not know if the Universe 
has a finite size. Second, doing finite time experiments we have to deal 
with some horizon and that does not depend on a size of the Universe. 
It is unclear how the cut off due to the horizon problem will interfere 
with the expansion of the Universe and its radius (if $R_U$ is finite).
\end{itemize}
The discussed effects are small and not detectable now. It is even not 
clear whether they may be detected in principle, however, they demonstrate 
a clear indication that
\begin{itemize}
\item a property of fundamental basic particles, like their charge and 
mass of the electron, should vary with time;
\item a property of compound objects, such as atoms and nucleus, 
should vary with time even if properties of their components are not varying.
\end{itemize}
The main question is the following: is there any reason for a faster 
variation, which can be detected with current experimental accuracy? 
This question has not yet been answered.

\section{Time and space variations}

Most considerations in literature have been devoted to the time 
variation. However, an astrophysical search (which has only provided us 
with possibly positive results) cannot distinguish between space and 
time variations, since remote astrophysical objects are separated from us 
both in time and space. 

To accept space variation is perhaps essentially the same as to suggest 
existence of some Goldstone modes. There is none for the Standard Model 
of the electroweak interactions, there are some experimental 
limitations on the Goldstone modes for Grand Unification Theories 
(see e.g. \cite{sgk:rpd}), but it is difficult to exclude them completely. 
Another option is some domain structure. In the case of ``large'' 
domains with the finite speed of light and horizon any easy conjunction 
of two domains is unlikely even reducing the total vacuum energy. 
A domain structure can be formed at the time of inflation when the 
Universe was expanding so fast that in a very short time two 
previously causality-connected points could be very far from each  
other -- out of horizon of each other. There is a number of reasons 
that a domain structure due to a parameter directly related to the 
vacuum energy cannot exist, since the energy would tend to reach its 
minimum. But if a construction like the Cabibbo-Kobayashi-Maskawa (CKM) 
matrix is a result of spontaneously broken symmetry, we could expect 
some minor fluctuations of CKM parameters, such as the Cabibbo angle, which 
were approximately, but not exactly, the same at some early time with 
their evolution being completely independent because of the horizon 
problem. CKM contributions are due to the weak interactions for hadrons 
and they slightly shift magnetic moments of proton and neutron 
at a fractional level of $10^{-5}$ and that is how such effects could be 
studied via precision spectroscopy. They are also important for the 
neutron lifetime and their variation could change the nuclear 
synthesis phemonena. We also have to underline that the space 
distribution with an expansion of the horizon and on their way to 
an equilibrium should provide some time evolution.

\section{Scenario and hierarchy}

A possibility of time variation of the values of the fundamental 
constants at a cosmological scale was first suggested quite a long time 
ago \cite{sgk:dirac,sgk:dyson}, but we still have no basic common 
idea on a possible nature of such a variation. A number of papers 
were devoted to the estimation of a possible variation of one of 
the fundamental constants (e.g. the fine structure constant $\alpha$) 
while a possible variation of any other properties is neglected. As we 
stated in \cite{sgk:cjp}, one has to consider a variation of all 
constants with approximately the same rate. However, some hierarchy 
(with rates different by an order of magnitude or even more) can be 
presented and it strongly depend on a scenario. There is a number 
of ``model dependent'' and ``nearly model independent'' estimations 
of the variation of the constants and their hierarchy.
\begin{itemize}
\item Any estimation based on running constants in SU(5) or in a similar 
unification theory is rather ``near model independent''. 
In particular, that is related to a statement on a faster variation of 
$m_p/M$ than $\alpha$ (see e.g. 
\cite{sgk:marciano,sgk:fritzsch,sgk:Langacker}).
\item Any estimation in the Higgs sector of SU(5) and other GUTs 
\cite{sgk:Langacker}, SUSY, quantum gravitation, strings {\em etc} 
strongly depends on the model.
\end{itemize}
We would like to clarify what is model-dependent in 
``near model independent'' considerations. It does not strongly depend 
on model suggestions in particle physics, but one still needs a basic 
suggestion on why (and how) any variation can happen. There may be a 
universal cause  all the time, or there may be a few ``phases'' 
with different causes dominating at different stages {\em etc}. 
What could be a basic cause for the dynamics? E.g. the basic 
suggestion for an SU(5) estimation is that everything can be derived 
from the Lagrangian \cite{sgk:marciano,sgk:fritzsch,sgk:Langacker} with 
varying parameters. In other words, for some reason there is 
dynamics operating within the Lagrangian.
\begin{itemize}
\item A supporting example is a multidimensional Universe with 
compactification of extra dimensions and the compactification 
radius $R$ as an ultraviolet cut-off $\Lambda=\hbar/Rc$ 
(see e.g. \cite{sgk:marciano}). Slow variation of $R$ is suggested 
(e.g. an oscillation at a cosmological time scale). All variations 
of the constants arise from the basic Lagrangian via the 
renormalization with a variation of the cut off and a variation in the 
Higgs sector induced   directly by the variation of $R$. 
\item On the contrary, it may be suggested that dynamics comes from 
a quantum nature of space-time and in terms of the Lagrangian 
that could lead to some new effective terms violating some basic 
symmetries of the ``unperturbed'' Lagrangian (indeed as a 
small perturbation). In such a case no reason due to SU(5) is 
valid and one has to start with a description of the perturbing terms.
\end{itemize}
Both options are open.

The ``model dependent'' estimations involve more unknown factors, 
they need understanding of both: a unification/extension scheme and 
a cause for the variation.

We need to mention an option that in principle the fundamental 
constants might be calculable. That does not contradict their 
variations, which can be caused by presence of some amount of matter, 
or by an oscillation of the compactification radius {\em etc}. In such 
a case, the truly fundamental constants $\alpha_0\sim 10^{-2}$ 
(the bare electric charge), $m_e^0/M_P\sim 10^{-22}$, 
$\Lambda_{\em QCD}/M_P\sim 10^{-20}$ are of very different order 
of magnitude (here $M_P$ is the Planck mass). The constants 
($\alpha$ and $(m, \Lambda)/M_P$) of so different order of magnitude 
can be either coupled logarithmically or not coupled at all. 
In the case of $\alpha$ and $\Lambda_{\em QCD}/M$ there is 
some understanding of this logarithmic coupling 
(see e.g. \cite{sgk:marciano,sgk:fritzsch}) which is mainly model 
independent (a model dependent part is a relation between $M_P$ and 
a mass of Grand Unification Theory $M$ which enters  relationships 
between the constants). In the case of $m_e^0/M_P$ model dependence 
is essential. However, as it is explained above, it is difficult 
to realize if any approximate relations between the constants 
are helpful or not. A crucial question is whether the variation 
supports the relations between the constants or violates them.

\section{Atomic and molecular spectroscopy and 
fundamental constants \label{sgk:trans}}

There are three most accurate results on a possible variation of the 
constants achieved recently. One of them is related to the Oklo 
fossil nuclear reactor \cite{sgk:oklo} and a position of 
the samarium resonance (\ref{sgk:149}). The result is negative 
and the assigned variation rate for the fundamental constants 
varies between $10^{-17}$ and $10^{-19}$ yr$^{-1}$ 
\cite{sgk:shlyakhter,sgk:irvine,sgk:damour,sgk:fujii}. However, 
the interpretation is rather unclear because there is no reliable way 
of studying the position of the resonance in terms of the 
fundamental constants. 

Two other results are related to spectroscopy:
\begin{itemize}
\item A study of the absorption spectra of some quasars led to a 
positive result on a variation of the fine structure constant of a 
part in $10^{15}$ per a year at $6\,\sigma$ level \cite{sgk:webb02} 
(see also earlier papers on a 4$\,\sigma$ positive result \cite{sgk:Flam}). 
Meanwhile, a search for a variation of $m_e/m_p$ showed a variation 
at a fractional level of $(5\pm2)\times 10^{-15}$~yr$^{-1}$ \cite{sgk:ivanchik}.
\item A comparison of hyperfine intervals for the ground state in 
cesium-133 and rubidium-87 shows no variation of the ratio 
of their frequencies at a level a part in $10^{15}$ \cite{sgk:Rb}. 
The ratio of these frequencies is more sensitive to a variation 
of $g_p$ than $\alpha$ \cite{sgk:cjp}. 
\end{itemize} 
Because of importance of the spectroscopic data, we briefly discuss the 
behavior of the frequency of different kinds of transitions as a function 
of the fundamental constants.

Any transition frequency can be presented in the form
\be
f = f_{\rm NR} \times F_{\rm Rel}(\alpha)\;,
\ee
where $f_{\rm NR}$ is the frequency in the leading non-relativistic 
approximation and $F_{\rm Rel}(\alpha)$ is the relativistic correcting 
factor. Scaling behavior of the non-relativistic results is summarized 
in Tables \ref{sgk:t:at} and \ref{sgk:t:mol}. The relativistic correction 
is a result of perturbative calculation of some singular terms 
since the relativistic effects are enhanced at short distances 
equivalent to the large momentum transfer. In neutral atoms and ions 
with only a few electrons stripped, the electron is located in the 
Coulomb field with a low effective charge of the screened nucleus 
at a long distance (e.g. $Z_{\rm eff}\simeq 1$ for neutral 
alkali atom). On the contrary, at a short distance the electron 
interacts rather with the bare nucleus and the effective charge 
is close to the nuclear charge $Z$. As a result, the correcting 
factor behaves as
\be
F_{\rm Rel}(\alpha)= 1 +C_2(Z\alpha)^2 + ...\;,
\ee
and at high $Z$ (e.g. for ytterbium and mercury) the correction is 
not small any more (see e.g. \cite{sgk:prestage,sgk:dzuba}).

\begin{table}[t]
\caption{Scaling behavior of atomic transitions. $\mu$ is the nuclear 
magnetic moment. References are given to the papers where the 
scaling behavior was first discussed. Importance of the 
relativistic corrections for the hyperfine structure was first 
understood in \protect\cite{sgk:prestage}, while for other 
transitions it was discussed in \protect\cite{sgk:dzuba}.\label{sgk:t:at}}
\vspace{0.2cm}
\begin{center}
\footnotesize
\begin{tabular}{|c|c|c|}
\hline
Transition & Energy scaling & Refs.\\
\hline
Gross structure & $Ry$ & \protect\cite{sgk:savedoff}\\
\hline
Fine structure & $\alpha^2Ry$ & \protect\cite{sgk:savedoff}\\
\hline
Hyperfine structure & $\alpha^2(\mu/\mu_B)Ry$ & \protect\cite{sgk:savedoff}\\
\hline
Relativistic corrections & Extra $\alpha^2$ & \protect\cite{sgk:prestage,sgk:dzuba}\\
\hline
\end{tabular}
\end{center}
\end{table}

\begin{table}[t]
\caption{Scaling behavior of molecular transitions. 
It is assumed that $m_p=m_n$ and the nuclear mass is $A\times m_p$. 
References are given to the paper where the scaling behavior was first discussed.\label{sgk:t:mol}}
\vspace{0.2cm}
\begin{center}
\footnotesize
\begin{tabular}{|c|c|c|}
\hline
Transition & Energy scaling & Refs.\\
\hline
Electronic structure & $Ry$ & \protect\cite{sgk:thompson}\\
\hline
Vibration structure & $(m_e/m_p)^{1/2}Ry$ & \protect\cite{sgk:thompson}\\
\hline
Rotational structure & $(m_e/m_p)Ry$ & \protect\cite{sgk:thompson}\\
\hline
\end{tabular}
\end{center}
\end{table}

Different scaling behavior of the non-relativistic transition 
frequencies allows to perform efficient comparison to search for a 
possible variation of the fundamental constants. The most accurate 
astrophysical results were obtained studying transitions of 
the same type \cite{sgk:webb02,sgk:Flam}, but with essentially 
different values of the nuclear charge $Z$ and thus with 
different relativistic corrections \cite{sgk:dzuba}. 

\section{Hyperfine structure and nuclear magnetic moments}

Looking for a variation of the fundamental constants with the 
help of the hyperfine structure, one needs to deal with the nuclear 
magnetic moments. There is no accurate model which allows to present 
the nuclear magnetic moments in terms of the fundamental constants. 
The only available model, the Schmidt model, is not really accurate. 
We summarize in Table \ref{sgk:t:hfs} the magnetic moments derived 
from the Schmidt model in comparison with the actual values for the 
atoms applied for the frequency standards (see also 
\cite{sgk:cjp}). The Table contains also data on relativistic corrections. 

\begin{table}[t]
\caption{Nuclear magnetic moments $\mu$, nuclear structure effects 
and relativistic effects for the atoms involved in precision microwave 
measurements. The uncertainty of the calculation in 
\protect\cite{sgk:prestage,sgk:casimir} is estimated by comparing results 
on cesium and mercury in \protect\cite{sgk:prestage,sgk:casimir} and 
\protect\cite{sgk:dzuba}. The actual values of the nuclear 
magnetic moments are taken from \protect\cite{sgk:firestone}.
\label{sgk:t:hfs}}
\vspace{0.2cm}
\begin{center}
\footnotesize
\begin{tabular}{ |c|c|c|c|c|c|}
\hline
$Z$ &Atom & Schmidt value   & Actual value  & Relativistic        & Sensitivity \\
    &     & for $\mu$   & for $\mu$ & factor              & to $\alpha$ variation\\
    &     & ($\mu_S/\mu_N$) & ($\mu/\mu_S$) & $F_{\rm rel}(\alpha)$ & $\partial \ln\big(F_{\rm rel}(\alpha)\big)/\partial \ln\alpha$ \\
\hline
1&H              & $g_p/2$        & 1.00 & 1.00 & 0.00\\
4&$^9$Be$^+$     & $g_n/2$        & 0.62 & 1.00 & 0.00\\
37&$^{85}$Rb      & $5/14(8-g_p)$ & 1.57 & 1.15, \cite{sgk:prestage,sgk:casimir} & 0.30(6), \cite{sgk:prestage,sgk:casimir} \\
37&$^{87}$Rb      & $g_p/2+1$      & 0.74 & 1.15, \cite{sgk:prestage,sgk:casimir} & 0.30(6), , \cite{sgk:prestage,sgk:casimir}  \\
55&$^{133}$Cs     & $7/18(10-g_p)$ & 1.50 & 1.39, \protect\cite{sgk:dzuba} & 0.83, \protect\cite{sgk:dzuba} \\
70&$^{171}$Yb$^+$ & $-g_n/6$       & 0.77 & 1.78 & 1.42(15), \cite{sgk:prestage,sgk:casimir} \\
80&$^{199}$Hg$^+$ & $-g_n/6$       & 0.80 & 2.26, \protect\cite{sgk:dzuba} & 2.30, \protect\cite{sgk:dzuba}\\
\hline
\end{tabular}
\end{center}
\end{table}

One can see that nuclear effects, responsible for a correction to 
the Schmidt model, are comparable to the relativistic effects, 
responsible for atomic corrections. 
Note the significant corrections to the Schmidt model for 
cesium-133 and rubidium-85. They are large because of a destructive 
interference of spin and orbit contributions, an essential 
cancellation of the leading term enhancing 
the corrections. The primary frequency standards are based 
on the hyperfine interval in cesium and the large corrections 
to the Schmidt value of the nuclear magnetic moment of 
cesium-133 are annoying for a direct interpretation of any 
absolute measurement, which is actually a comparison of some transition 
with the cesium standards.

\section{Optical transitions \label{sgk:opt}}

The essential nuclear effects related to the nuclear 
magnetic moment lead to a problem of a reliable interpretation of 
the data. Much more reliable results are delivered by studying  pure 
optical transitions \cite{sgk:1s2s,sgk:Ca,sgk:In,sgk:Yb,sgk:Hg,sgk:Hgnew} 
which can be obtained via a direct comparison of two optical 
frequencies, or indirectly via independent absolute measurements 
of those frequencies in units determined by the cesium microwave 
transition. Both kinds of comparison are now available for the 
frequency metrology after a development of the new frequency 
chain based on the so-called frequency comb 
\cite{sgk:chain}. The most accurate data are summarized in 
Table ~\ref{sgk:t:opt}.

\begin{table}[t]
\caption{Optical transitions: most accurate results and 
sensitivity of the optical transitions to a time variation 
of $\alpha$.\label{sgk:t:opt}}
\vspace{0.2cm}
\begin{center}
\footnotesize
\begin{tabular}{|c|c|c|c|c|}
\hline
$Z$&Atom & Frequency & Fractional & Sensitivity to $\alpha$ variation \\
&& [Hz] &uncertainty &$\partial \ln\big(F_{\rm rel}(\alpha)\big)/\partial \ln\alpha$,\protect\cite{sgk:dzuba}\\
\hline
1&H      & 2\,466\,061\,413\,187\,103(46), \protect\cite{sgk:1s2s}& $2\times 10^{-14}$ & 0.00 \\
20&Ca     & 455\, 986\,240\,494\,158(26), \protect\cite{sgk:Hg}& $6\times 10^{-14}$ & 0.03\\
49&In$^+$ & 1\,267\,402\,452\,899\,920(230), \protect\cite{sgk:In}& $18\times 10^{-14}$ & 0.21 \\
70&Yb$^+$ & 688\,358\,979\,309\,312(6), \protect\cite{sgk:Yb}& $0.9\times 10^{-14}$ &1.03 \\
80&Hg$^+$ & 1\,064\,721\,609\,899\,143(10), \protect\cite{sgk:Hg}& $0.9\times 10^{-14}$ & - 3.18\\
\hline
\end{tabular}
\end{center}
\end{table}

\begin{figure}
\centerline{\epsfig{figure=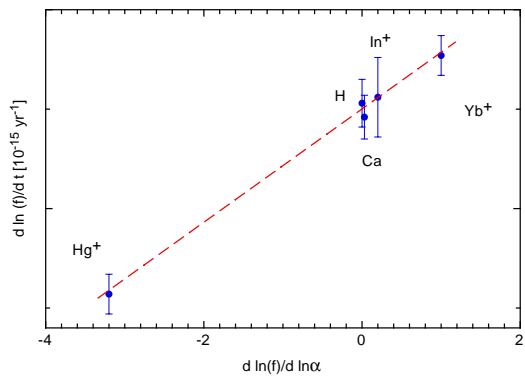,height=2.0in}}
\caption{An expected signature of the time variation of the 
fine structure constant.\label{sgk:fig:an}}
\end{figure}

An important feature of the optical transitions related to the 
gross structure is that they can be described with the help of two 
constants only: the Rydberg constant and the fine structure constant. 
As a result, a time variation of any frequency can be presented 
in the form
\be
\frac{\partial \ln{f}}{\partial t} = A + B \frac{\partial \ln{F_{\rm rel}(\alpha)}}{\partial \ln\alpha}\;,
\ee
where
\be\nonumber
A = \frac{\partial \ln\bigl(Ry\bigr)}{\partial t}~~~~~~~~~~~~~~~{\rm and}~~~~~~~~~~~~~~~
B = \frac{\partial \ln\alpha}{\partial t}\;.
\ee
While a variation of the Rydberg constant as we discussed above 
could have no simple interpretation in terms of the 
fundamental constants, a time variation of the fine structure constant 
would have a direct and simple interpretation. An expected 
signature of the time variation of $\alpha$ is depicted in Fig.~\ref{sgk:fig:an}. In near future five accurate results are expected. Three of them are related to ``near $\alpha$-independent'' results (hydrogen, calcium, indium) and they should play a role of an anchor. Two others (mercury and ytterbium) are strongly $\alpha$ dependent and the dependence is significantly different (see Table ~\ref{sgk:t:opt}).

\section{Current laboratory limitations}

Current laboratory limitations are summarized in 
Table ~\ref{sgk:t:cur}. A limitation on the time variation of the 
proton $g$ factor is derived assuming that the nuclear corrections 
to the Schmidt model are not important. That indeed cannot be considered 
as a reliable approach. All other limitations are obtained 
in a more reliable way. As was pointed out in \cite{sgk:cjp} 
(see also Table \ref{sgk:t:hfs}), 
the hyperfine interval in cesium is very difficult for interpretation 
because of significant nuclear corrections to the Schmidt model. 
Fortunately, it was demonstrated that there is no variation on a 
level of a part in $10^{15}$ per a year for a ratio of 
cesium-to-rubidium hyperfine structure \cite{sgk:Rb} 
(as a matter of fact that is the strongest laboratory limitation 
on a variation of a transition frequency). The hyperfine interval 
in the ground state of the rubidium-87 in contrast to cesium-133 
can be sufficiently well described by its non-relativistic part with 
use of the Schmidt model for the nuclear magnetic moment 
\cite{sgk:cjp} (see also Table \ref{sgk:t:hfs}).

\begin{table}[t]
\caption{Current limitations on a possible time variation of the 
fundamental constants. \label{sgk:t:cur}}
\vspace{0.2cm}
\begin{center}
\footnotesize
\begin{tabular}{|c|c|}
\hline
Fundamental & Limitation for\\
constant & variation rate\\ \hline
$\alpha$ & $1\times 10^{-14}$ yr$^{-1}$ \\ \hline
$m_e/m_p$ &$2\times 10^{-13}$ yr$^{-1}$ \\ \hline
$\alpha^2 \mu_p/mu_e$ & $6\times 10^{-14}$ yr$^{-1}$ \\ \hline
$\alpha^{-3} \mu_p/mu_e$ & $7\times 10^{-15}$ yr$^{-1}$ \\ \hline
$\mu_p/mu_e$ & $2\times 10^{-14}$ yr$^{-1}$ \\ \hline
$g_p$ & $4\times 10^{-16}$ yr$^{-1}$ \\ \hline
$\alpha^2 \mu_n/mu_e$ & $8\times 10^{-14}$ yr$^{-1}$ \\ \hline
$\mu_n/mu_e$ & $6\times 10^{-14}$ yr$^{-1}$ \\ \hline
$g_n/g_p$ & $5\times 10^{-14}$ yr$^{-1}$ \\ \hline
\end{tabular}
\end{center}
\end{table}

\section{Precision spectroscopy: tests and reliability}

Recent progress in frequency metrology delivered us a number 
of results, essentially more accurate than any previous data and the 
expected results can be even more accurate. In such a case 
we need to be sure that the results are reliable. 
In this section we briefly discuss possible tests of the 
accurate frequency measurements.
\begin{itemize}
\item The cesium hyperfine interval plays a special role in 
physics because of the definition of the second. It is realized 
in a number of laboratories and a comparison of 
different cesium standards is an important metrological work. 
The comparison shows that we have a sufficient understanding of the 
accuracy of cesium experiments (see e.g. \cite{sgk:cs}).
\item Study of the ${}^3P_1-{}^1S_0$ transition in neutral calcium 
were performed independently at NIST \cite{sgk:Hg} and PTB 
\cite{sgk:Ca} and the results are consistent.
\item Hyperfine structure of the ground state of ytterbium ion 
was measured independently at PTB and NML \cite{sgk:yb}. The results 
are consistent.
\item An important approach to test systematic sources part by part 
may be a measurement of the isotopic shift and its comparison with theory. 
If theory is not accurate enough, there is still an option for a 
precision study. Theory is helpful to fix the form of dependence 
on the nuclear mass and the nuclear charge radius and  
the shape of the dependence may be checked via fitting.
\item Similar test can be performed studying the hyperfine structure. 
E.g. a comparison of the $1s-2s$ transitions in hydrogen for 
different spin states \cite{sgk:2s}. Since the hyperfine splitting 
in the ground state is known with a high accuracy \cite{sgk:1s,sgk:cjp}, 
the comparison of the $1s-2s$ ultraviolet transitions yields us 
a value of the hyperfine interval in a metastable $2s$ state. 
The result is more accurate that one directly derived from a 
microwave measurement \cite{sgk:hessels} and in good agreement 
with theory \cite{sgk:2stheo}. The transitions under study 
\cite{sgk:2s} as well as comparison with theory and early microwave 
measurements are summarized in Fig.~\ref{sgk:fig:2s}.

\begin{figure}
\centerline{
\epsfig{figure=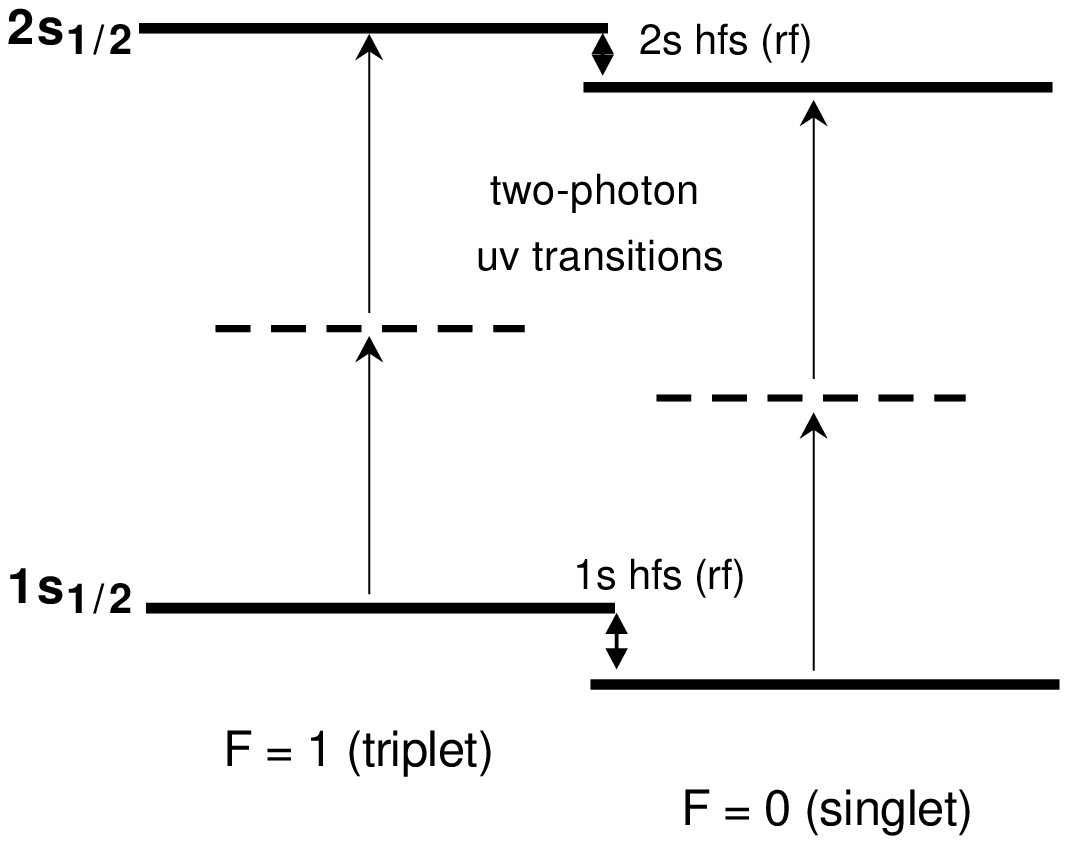,height=1.7in}~~~~~~~
\epsfig{figure=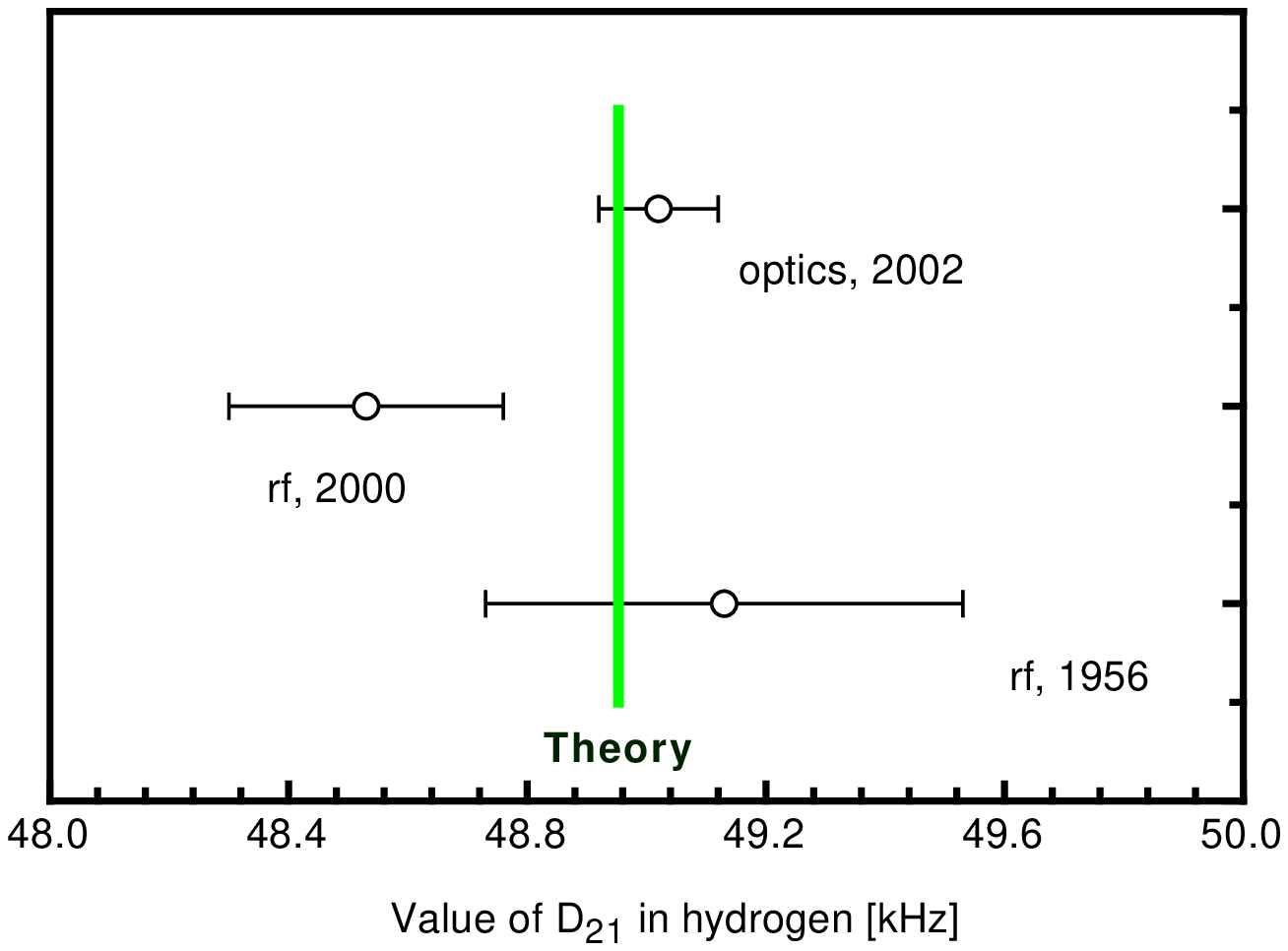,height=1.7in}}
\caption{Hyperfine structure in the hydrogen atom: levels scheme of an optical measurement of the $2s$ {\rm hfs} interval and a comparison of theory to experiment for 
$D_{21}= 8\times \nu_{\rm HFS}(2s)-\nu_{\rm HFS}(1s)$.\label{sgk:fig:2s}}
\end{figure}
\end{itemize}

\section{Summary \label{sgk:sum}}

A comparison of different kinds of search for a possible time and space 
variations of the fundamental constants is summarized in Table~\ref{sgk:t:com}. The characteristic level of the limitations is given suggesting a linear time dependence. In the case of oscillation the limitation from geochemical search and from astrophysical observations should be weakened by a factor $2\Delta t/T$, since the period of oscillation $T$ can be shorter than the time separation $\Delta t$. We note that in the case of laboratory limitation the results should depend on a current phase of oscillations. Another problem with interpretation of the astrophysical data is a separatation of space and time variations. The different kinds of search offer access to different sets of constants and their reliability depends on whether they are affected by the strong interactions.

\begin{table}[t]
\caption{Comparison of different kind of search for a possible time and space variation of the fundamental constants. \label{sgk:t:com}}
\vspace{0.2cm}
\begin{center}
\footnotesize
\begin{tabular}{|c|c|c|c|c|}
\hline
&Geochemistry & Astrophysics & Laboratory & Laboratory \\
&&&&(optics) \\
\hline
Drift or oscillation & $\Delta t \sim 10^9$ yr & $\Delta t \sim 10^9$ yr & $\Delta t \sim 1-30$ yr & $\Delta t \sim 1-10$ yr \\ \hline
Space variations & $\Delta l \simeq 10^9~c\times$yr & $\Delta l \simeq 10^{10}~c\times$yr & 0 & 0 \\ \hline
Level of limitations & $10^{-17}$ yr$^{-1}$ & $10^{-15}$ yr$^{-1}$ & $10^{-15}$ yr$^{-1}$ & $10^{-14}$ yr$^{-1}$ \\ \hline
Present results & Negative & Positive ($\alpha$) & Negative & Negative \\ \hline
Variation of $\alpha$ & not reliable & accessible & accessible & accessible \\ \hline
Variation of $m_e/m_p$ & not accessible & accessible & accessible & not accessible \\ \hline
Variation of $g_p$ & not accessible & accessible & accessible & not accessible \\ \hline
Variation of $g_n$ & not accessible & not accessible & accessible & not accessible \\ \hline
Strong interactions & not sensitive & not sensitive & sensitive & not sensitive \\ \hline
\end{tabular}
\end{center}
\end{table}

Despite a number of advantages and disadvantages of different approaches 
there is no favorite way. Since we have no background theory, 
we need to try as many searches as possible and as different as possible.

There are a number of problems which may be of interest and we'd 
like to attract attention to a few of them.
\begin{itemize}
\item A comparison of hyperfine intervals in the ground state of $^{85}$Rb 
and $^{87}$Rb allows to remove any variation of the fine structure 
constant due to atomic interactions and the frequency ratio is sensitive 
only to the proton $g$ factor via the Schmidt model and to strong 
interactions via corrections to the Schmidt model. Separation 
of atomic and nuclear physics can be helpful as a test measurement 
when a number of microwave intervals related to the hyperfine 
structure will be studied.
\item Actual nuclear magnetic moments of $^{199}$Hg 
and $^{171}$Yb are very close (the difference is below 5\%) and their 
Schmidt values are the same (see Table \ref{sgk:t:hfs}). If that is a 
systematic effect, a comparison of the hyperfine intervals in 
these two ions can give a reliable result on a possible variation 
of the fine structure constant $\alpha$. We need better understanding 
of the magnetic moments of $^{199}$Hg and $^{171}$Yb.
\item Discussing different approaches, we need to mention an idea 
of \cite{sgk:peik} to study a 3.5 eV nuclear transition in $^{229}$Th 
which lies in the optical domain. Its comparison with atomic 
transitions can have indeed no reliable interpretation, but 
the nuclear transition is very different from  atomic transitions and may 
be sensitive to effects not detectable with other methods.
\item Another approach related to the nuclear properties 
suggested \cite{sgk:cjp} precision studies of the nuclear magnetic 
moment with extremely small values, which are expected to be very 
sensitive to detuning of the fundamental constants. Indeed, it is not 
possible to measure a nuclear magnetic moment accurately enough. 
However, looking for their variations one can study the 
hyperfine structure of proper ions. As an example of an extremely 
small magnetic moment, let us mention ${}^{198}_{81}$Tl with a 
magnetic moment below $10^{-3}\;\mu_N$ ($T_{1/2}=5.3(5)$ h), 
$^{153}_{~62}$Sm ($\mu = - 0.022\,\mu_{N}$, $T_{1/2}=46$ h) 
and $^{192}_{~79}$Au ($\mu = - 0.009(2)\,\mu_{N}$, 
$T_{1/2}=4.9$ h) \cite{sgk:firestone}. Understanding the 
nature of such small values is also necessary.
\item One more question related to the subject: 
can we detect the expansion of the Universe in some laboratory 
experiments? The expansion leads to the red shift of the photons at 
a level of $10^{-10}$ yr$^{-1}$, however, there is no way to study 
a photon emitted a year ago in laboratory experiments. 
A chance can appear if we can use objects (planets, spacecrafts) at 
a distance related to $c\times (1-10)\;$min.
\end{itemize}

A search of a possible variation of the values of the fundamental 
constant presents a specific field involving both fundamental and 
applied physics. A search for new physics is based on  frequency metrology 
providing a high motivation. The frequency metrology 
presents now limitations which are somewhat weaker than those 
from astrophysics but it has showed significant progress last years 
and it seems that higher accuracy of the laboratory measurements is 
just a matter of time and new results will be coming soon.

\section*{Acknowledgments}
The author is grateful to V. Flambaum, H. Fritzsch, T. W. H\"ansch, J. L. Hall, M. Kramer, W. Marciano, M. Murphy, L. B. Okun, E. Peik, T. Udem, D. A. Varshalovich, M. J. Wouters and J. Ye for useful and stimulating discussions. The work has been supported in part by RFBR.


\end{document}